\documentclass{article}

\usepackage{PRIMEarxiv}

\usepackage[utf8]{inputenc} 
\usepackage[T1]{fontenc}    
\usepackage{hyperref}       
\usepackage{url}            
\usepackage{booktabs}       
\usepackage{amsfonts}       
\usepackage{nicefrac}       
\usepackage{microtype}      
\usepackage{lipsum}
\usepackage{amsthm}
\usepackage{fancyhdr}       
\usepackage{graphicx}       
\graphicspath{{media/}}     

\usepackage{amsmath,amssymb,amsfonts}
\usepackage{algorithmic}
\usepackage{graphicx}
\usepackage{textcomp}
\usepackage{multirow}
\usepackage{array}
\usepackage{subcaption}
\usepackage{amsmath,nccmath}
\usepackage{amsfonts}
\usepackage{tablefootnote}

\newcommand{\probP}{\text{I\kern-0.15em P}}

\newtheorem{remark}{Remark}

\usepackage{subcaption}
\usepackage{pgfplots}
\pgfplotsset{compat=1.18}
\usepackage{tikz}
\usetikzlibrary{decorations.pathreplacing}
\usetikzlibrary{shapes.geometric, arrows.meta, positioning, calc}

\title{Measuring Epistemic Unfairness for Algorithmic Decision-Making}

\author{
  Camilla Quaresmini \\
  Politecnico di Milano \\
  Milan, Italy\\
  \texttt{camilla.quaresmini@polimi.it} \\
  \And
   Lisa Piccinin \\
   Politecnico di Milano \\
   Milan, Italy \\
   \texttt{lisa.piccinin@polimi.it} \\
  \And
  Valentina Breschi \\
  Eindhoven University of Technology \\
  Eindhoven, The Netherlands\\
  \texttt{v.breschi@tue.nl} \\
}

\begin{document}
\maketitle

\begin{abstract}
Algorithmic systems increasingly function as epistemic infrastructures that govern the conditions of interpretative access and social belief. Yet, mainstream auditing strategies operationalize fairness primarily in predictive terms - error rates, calibration, or group-level parity - leaving epistemic harms under-theorized and under-measured. We propose a quantitative framework for evaluating forms of epistemic injustice in algorithmic environments. First, we introduce a deficit-based template that models epistemic injustices as gaps between ideal and realized conditions across features such as credibility, uptake, and epistemic agency. We map these deficits to concrete stages of algorithmic mediation, showing how epistemic injustice can persist even when standard fairness constraints are satisfied. Drawing on distributive fairness indices, we distinguish two evaluation stances: resource inequality, where indices are applied to distributions of epistemic goods directly, and capability/rights inequity, where indices are applied to output-induced epistemic opportunity. We provide an epistemic translation of canonical indices, illustrating how they diagnose complementary signatures of unfairness - such as exclusionary tails and hierarchical concentration - and support longitudinal auditing under iterative deployment. We also provide a simulation study of a recommender-mediated opinion dynamics setting, showing how the proposed indices capture the evolution of epistemic unfairness under repeated platform interventions.
The result is a measurement framework that makes the epistemic dimension of algorithmic harms explicit for system design and evaluation.
\end{abstract}


\section{Introduction}

The architecture of the contemporary information ecosystem is deeply governed by algorithmic infrastructures. Search and recommendation rankings, content-moderation pipelines, automated labeling, and reputation scores shape which claims are heard and how they are interpreted \cite{cobbe2021algorithmic,gillespie2014relevance,milano2025algorithmic,gerrard2020behind}. Contemporary platforms instantiate a ``scored society'' \cite{benjamin2019captivating}, where pervasive evaluation 
practices can reproduce existing inequalities and create new forms of marginalization.
This scoring logic extends to the \emph{epistemic} realm. Because algorithmic systems are intrinsically epistemic \cite{alvarado2023ai,symons2022epistemic} - they deal with the production, diffusion 
and organization of knowledge - they effectively assign users' credibility and the legitimacy of their knowledge claims, while settling classificatory boundaries that shape interpretation. As large-scale infrastructures \cite{weerts2021introduction}, they can systematically reshape 
trust relations, access to interpretative resources, and 
epistemic participation. When such systems privilege some groups' knowledge claims while discounting others, they can amplify and automate epistemic injustice \cite{stewart2022perfect,coeckelbergh2022political,coeckelbergh2025ai}.
Social epistemology offers a vocabulary for these harms under the label \emph{epistemic injustice}, that is, wrongs suffered by people as knowers \cite{fricker2007epistemic}. In particular, on Miranda Fricker's foundational account, \emph{testimonial injustice} occurs when prejudice leads hearers to assign a speaker less credibility than deserved. On the other hand, \emph{hermeneutical injustice} arises when structural gaps in interpretative resources prevent agents from understanding 
experiences. Recent literature argues that algorithmic systems can reproduce and intensify these injustices, for example by systematically discounting marginalized users' credibility, or entrenching hermeneutical gaps through classificatory schemes \cite{medvecky2018fairness,edenberg2023epistemic,nihei2022epistemic}.
However, despite the crucial epistemic character of contemporary platforms, the dominant 
discourse on algorithmic fairness rarely treats epistemic injustice as a fairness problem. This is partly because the literature has largely operationalized harms in predictive and distributive terms, focusing on error rates, calibration, and parity relations across 
demographics \cite{barocas2020fairness,mehrabi2021survey,pessach2022review}. These metrics are essential for many tasks, but they are not designed to capture epistemic goods such as credibility, interpretative resources, or epistemic agency, nor do they directly model how epistemic deficits can persist in algorithmic platforms. While some recent work begins to operationalize aspects of testimonial injustice in algorithmic settings (e.g., \cite{villa2025epistemic,quaresmini2025role}), the broader landscape of epistemic injustice documented in social epistemology remains under theorized and rarely measured in algorithmic fairness research. 
In this context, the paper aims to make epistemic injustice tractable for algorithmic system design and evaluation. Specifically, we do not seek to provide a comprehensive characterization of epistemic harms. Rather, we develop a diagnostic framework for characterizing certain aspects of epistemic injustice in algorithmic settings in distributive terms and measuring them over time. To do so, we proceed in two steps. First, we formalize epistemic injustice as a deficit in epistemic goods that 
algorithmic systems can induce or amplify, 
where the relevant ``good'' depends on the injustice type and application context. Second, we draw on a family of indices from economics, designed to measure uneven distributions of 
resources over time, to quantify the deficits. The deficit framework and distributive indices constitute two parts of the same diagnostic task. To investigate how they are patterned across individuals and groups,
we adopt a dual evaluation stance. First, we measure inequality in the distribution of epistemic goods.  Second, we measure inequity in the opportunities mediated by system outputs.
Unlike standard Machine Learning (ML) fairness metrics, which typically focus on error rates at single time, these indices capture the distributional structure of epistemic disadvantages across individuals and groups.
We thus translate philosophical accounts of epistemic injustice into index-based measures for evaluating epistemic unfairness in algorithmic systems. In this way, we frame \emph{epistemic fairness} as a mechanism-level lens for identifying how algorithmic systems generate, sustain, or amplify epistemic injustice.
We complement the framework with a simulation study, showing how interventions shape trajectories of epistemic inequality and output-level inequity over time.

The paper is structured as follows. Section \ref{sec:deficit} introduces a deficit-based template for epistemic injustice. Section \ref{sec:alg} maps the deficits to algorithmic systems, explaining why they may persist despite standard 
fairness constraints. Section \ref{sec:metrics} reviews classical fairness evaluations. Section \ref{sec:indices} introduces the two facets of epistemic auditing. Section \ref{sec:epistemic} presents and  translates classical fairness indices in epistemic terms.
Section \ref{sec:sim_example} illustrates how fairness indices can be used to evaluate the evolution of epistemic quantities through a simulative example. Section \ref{sec:conc} concludes with directions for future work.

\section{A Deficit-based Formal Treatment of Epistemic Injustice}\label{sec:deficit}
Epistemic injustice is not confined to interpersonal exchanges. In algorithmic environments, credibility, uptake, interpretative access, and epistemic participation are routinely mediated by ranking, moderation and labeling infrastructures. These mechanisms can systematically privilege some agents' contributions while discounting others, or stabilize interpretative gaps by enforcing dominant classificatory schemes. Epistemic harms can thus arise as properties of algorithmic systems behavior at scale. To investigate these harms in ways that inform the design of \emph{fair} algorithmic systems, we introduce a formal template that models epistemic injustice via deficits in epistemic goods.

\begin{figure}[]
    \centering
    \scalebox{.75}{\begin{tikzpicture}[
        node distance=1.2cm and 1.2cm,
        auto,
        block/.style={
            rectangle, draw, fill=blue!5,
            text width=3cm, text centered,
            rounded corners, minimum height=1.4cm,
            font=\small,
            align=center
        },
        system/.style={
            trapezium, draw, fill=gray!10,
            trapezium left angle=75, trapezium right angle=75,
            text width=2.2cm, text centered,
            minimum height=1.4cm,
            font=\small,
            align=center
        },
        index/.style={
            ellipse, draw, fill=green!5,
            text width=1.6cm,
            text centered,
            font=\scriptsize,
            minimum height=0.8cm,
            align=center
        },
        line/.style={draw, -{Stealth[scale=1.1]}, thick},
        dashline/.style={draw, -{Stealth[scale=1.1]}, thick, dashed, color=gray!80}
    ]

\node [block] (ideal) {%
\begin{tabular}{@{}c@{}}
\textbf{Epistemic Ideal ($I^*$)}\\[0.2ex]
{\scriptsize\itshape Credibility}\\
{\scriptsize\itshape Hermeneutical Access}\\
{\scriptsize\itshape Epistemic Agency}
\end{tabular}
};

\node [system, right=of ideal] (algo) {%
\begin{tabular}{@{}c@{}}
\textbf{Algorithmic}\\
\textbf{Mediation}\\[0.2ex]
{\scriptsize\itshape Ranking}\\
{\scriptsize\itshape Labeling}\\
{\scriptsize\itshape Moderation}
\end{tabular}
};

\node [block, right=of algo] (realized) {%
\begin{tabular}{@{}c@{}}
\textbf{Realized Condition ($I$)}\\[0.2ex]
{\scriptsize\itshape Effective Visibility}\\
{\scriptsize\itshape Uptake}\\
{\scriptsize\itshape Interpretive Opportunity}
\end{tabular}
};

\node [above=0.2cm of ideal, font=\scshape\tiny, color=gray!70] {Theory};
\node [above=0.2cm of algo, font=\scshape\tiny, color=gray!70] {Infrastructure};
\node [above=0.2cm of realized, font=\scshape\tiny, color=gray!70] {Measurement};

\draw [line] (ideal) -- (algo);
\draw [line] (algo) -- (realized);

\draw [decorate, decoration={brace, mirror, amplitude=8pt, raise=10pt}, thick] 
    (ideal.south west) -- (realized.south east) 
    node [midway, below=18pt, font=\small\bfseries] 
    {$\Delta_I = |I^* - I|$};

\node [index, right=1.4cm of realized, yshift=0.6cm] (resource) {Resource Inequality};
\node [index, below=0.5cm of resource] (capability) {Capability Inequity};

\draw [dashline] (realized.east) -- (resource.west);
\draw [dashline] (realized.east) -- (capability.west);

\end{tikzpicture}}
\caption{\textbf{The proposed framework for evaluating algorithmic epistemic injustice.}
The gap between the normative Epistemic Ideal ($I^*$) and the Realized Condition ($I$) is modeled as a measurable deficit ($\Delta_I$).}
\label{fig:framework}
\end{figure}
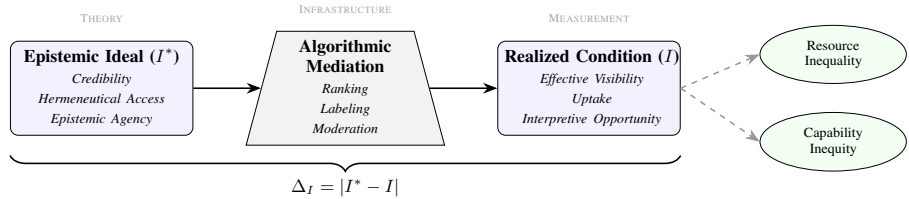

As illustrated in Figure \ref{fig:framework}, our approach treats algorithmic systems as mediating infrastructures that transforms epistemic ideals into realized outcomes. Specifically, let $\mathcal{I}$ be a set of epistemically relevant features (e.g., credibility, interpretative resources, epistemic agency). For an agent (or group) and context of interest, let $I^{\star}$ denote the \emph{ideal} epistemic condition and $I$ the corresponding \emph{actual} condition. The associated \emph{epistemic deficit} is
\begin{equation}\label{eq:deficit}
  \Delta_I
  \;=\;
  \bigl|I^{\star} - I\bigr|
  \;\ge\; 0 .
\end{equation}
Accordingly, $\Delta_{\mathcal{I}}$ is the magnitude of the gap between what is epistemically warranted and what is effectively realized.
This deficit 
framing is familiar in algorithmic fairness literature. Metrics such as statistical parity, equality of opportunity, and equalized odds quantify deviations from an ideal constraint. When used as soft constraints (e.g., as regularizers or loss terms), they effectively act as costs of unfairness, increasing as the deficit between ideal and actual behavior grows.

When epistemic deficits are systematic and structured by identity-based prejudices, they constitute epistemic injustice. Different forms of epistemic injustice correspond to different instantiations of $I^{\star}$ and $I$. For example, in testimonial injustice $I^{\star}$ can be taken as a subject's reliability and $I$ as the credibility attributed to them. In hermeneutical injustice, $I^{\star}$ can be taken as the interpretive resources available in principle and $I$ as those accessible in practice.

\begin{remark}
    Our deficit schema above treats shortfalls relative to an epistemic ideal as the primary measurable harm. Testimonial injustice, however, is often discussed as having a deficit-excess structure, i.e., credibility deficits for some speakers can be mirrored by credibility excesses for others \cite{fricker2007epistemic,coady2017epistemic}. In the context of algorithmic ranking and recommendation, this relationship is frequently causal rather than merely symmetrical. Because visibility and attention are scarce and competitively allocated, an unearned credibility excess for dominant users can directly generate a credibility deficit for marginalized ones. While the current framework prioritizes measuring deficits as a main indicator of harm, explicitly modeling the relational dynamics of epistemic excesses is an important direction for future work.
\end{remark}

\indent\textit{Modeling stance (distributive operationalization).}
Treating epistemic injustice as a deficit mechanism is a deliberate simplification adopted for measurement and evaluation in algorithmic settings. Indeed, our claim is not that epistemic injustice is exhaustively reducible to utility, welfare, or any distributive quantity. Rather, we identify aspects of epistemic injustice that admit formal representation through measurable proxies aligned with
philosophical concepts.
Testimonial and hermeneutical injustice are often characterized as discriminatory wrongs in epistemic relations, not merely as maldistributions of goods \cite{fricker2007epistemic,fricker2013epistemic}. Our formalization does not claim to exhaust this normative structure. In particular, we acknowledge that epistemic injustice is first and foremost an epistemic wrong. This means that subjects are wronged in their capacity as knowers, via distortions governed by erroneous, non-epistemic reasons. As emphasized by \cite{symons2022epistemic}, this epistemic wrong can be conceptually prior to, and come apart from, downstream distributive or socio-economic outcomes. But this does not mean it lacks a distributional manifestation within the epistemic domain. In algorithmically mediated settings, repeated discriminatory epistemic judgments (who is credible, who is intelligible) accumulate into stable patterns of epistemic standing, uptake, and interpretive access. We therefore treat epistemic injustice as an \emph{epistemic-distributive problem}, that is, a case in which discriminatory epistemic mechanisms produce patterned maldistributions of epistemic resources. This distributive lens resonates with broader work on inequality and inequity aversion, according to which normative evaluations depend 
also on how advantages and disadvantages are distributed across agents. In particular, classic accounts emphasize sensitivity to interpersonal asymmetries and to relative standing within a distribution \cite{fehr1999theory,bolton2006inequality}, which supports our treatment of epistemic harms as distributionally structured deficits in epistemic goods and opportunities.
To make this distributive operationalization concrete, we need to specify what counts as the relevant epistemic good in a given setting. Social epistemology offers useful distinctions here. In particular, even if not sharply bounded in existing literature, we distinguish: 
\begin{itemize}
     \item \emph{Epistemic and hermeneutical resources} (shared concepts, credibility, interpretative tools) \cite{fricker2007epistemic,ferguson2025hermeneutical}. These include shared concepts for making sense of experiences such as online harassment or algorithmic discrimination, or public credibility markers, e.g., verification badges on social networks;
     \item \emph{Epistemological resources} (individual-level reasoning strategies) \cite{elby201013}. These contain epistemic competences such as assessing source reliability, interpreting conflicting evidence, or recognizing misleading content;
     \item \emph{Epistemic capital} (standing, authority, and credit) \cite{alasuutari2016organisations,davis2018epistemic}. It involves socially recognized epistemic standing, such as reputation coming from citations, institutional affiliation, or visibility on a platform.
\end{itemize}
Different forms of epistemic injustice can be understood as systematic distortions in the distribution or access to these resources \cite{edidin2025beyond,dotson2014conceptualizing}. Repeated credibility attributions, access decisions, and classificatory constraints accumulate into stable inequalities in credibility standing, uptake, and interpretative access. Modeling these patterns as deficits according to Eq.~\eqref{eq:deficit} yields measurable variables that can be tracked and compared across groups, time, and system configurations (as shown in Tables \ref{tab:ei_technical} and \ref{tab:ei_technical_forced}). This move is consistent with work that treats credibility and interpretative participation as distributable epistemic goods, arguing that epistemic injustices can be fruitfully analyzed in deficit-excess and allocation terms \cite{coady2017epistemic}.
\begin{table}
  \caption{\textbf{Technical representations of epistemic injustices with a direct ideal--actual shortfall formulation.} Arrow orientation indicates the direction of epistemic movement.}
  \label{tab:ei_technical}
  \begin{tabular}{p{0.19\linewidth} p{0.30\linewidth} p{0.31\linewidth} p{0.09\linewidth}}
    \toprule
    Injustice & Ideal condition & Non-ideal condition & Deficit \\
    \midrule

    Contributory injustice \cite{dotson2011tracking} &
    All marginalized interpretative resources $i_{m \to c}$ are taken up in the
    collective resource $i_c$: \ $i_{m \to c} = i_c$. &
    Not all marginalized resources are incorporated into $i_c$ by dominant hearers: $i_{m\to c} < i_c$. &
    $\bigl|i_c - i_{m\to c}\bigr|$ \\

    Formative epistemic injustice \cite{nikolaidis2021third} &
    Each agent has access to collective educational resources $e_c$: $e_{a\leftarrow c} = e_c$. &
    Agent $a$'s access to $e_c$ is limited by institution: $e_{a\leftarrow c} < e_c$. &
    $\bigl|e_c - e_{a\leftarrow c}\bigr|$ \\

    Hermeneutical injustice \cite{fricker2007epistemic} &
    All agents can access collective
    interpretative resources $i_c$: $i_{a\leftarrow c} = i_c$. &
    Agent $a$ is denied access: $i_{a\leftarrow c} < i_c$. &
    $\bigl|i_c - i_{a\leftarrow c}\bigr|$ \\

    Prediscursive epistemic injury \cite{lobb2018prediscursive} &
    All agents enjoy an ideal level of epistemic agency $a_c$: $a_{a\leftarrow c} = a_c$.  &
    Agent $a$'s effective 
    epistemic
    agency is reduced 
    : $a_{a\leftarrow c} < a_c$. &
    $\bigl|a_c - a_{a\leftarrow c}\bigr|$ \\

    Testimonial injustice
    \cite{fricker2007epistemic,quaresmini2025role} &
    Speaker $s$'s assigned credibility matches reliability:
    $c_{h\leftarrow s} = r_s$. &
    Hearer $h$ assigns reduced credibility to $s$: 
    $c_{h\leftarrow s} < r_s$. &
    $\bigl|r_s - c_{h\leftarrow s}\bigr|$ \\

    Testimonial quieting \cite{dotson2011tracking} &
    Audience $h$ fully takes up speaker $s$'s potential contribution: $u_{h\leftarrow s} = u_s$. &
    Audience $h$ collapses uptake of $s$'s contribution to
    $u_{h\leftarrow s} < u_s$. &
    $\bigl|u_s - u_{h\leftarrow s}\bigr|$ \\

    Testimonial smothering \cite{dotson2011tracking} &
    Speaker $s$ can safely share epistemic content $e_s$ to audience $h$:
    $e_{s\to h} = e_s$. &
    Speaker $s$ self-censors fearing hostile conditions
    , sharing less 
    $e_{s\to h} < e_s$. &
    $\bigl|e_s - e_{s\to h}\bigr|$ \\

    Willful hermeneutical ignorance \cite{pohlhaus2017varieties} &
    Hearers $h$ make full use of interpretative tools $i_c$:
    $i_{h\leftarrow c} = i_c$. &
    Hearer $h$ intentionally restricts their interpretative uptake $i_h$:
    $i_{h\leftarrow c} < i_c$. &
    $\bigl|i_c - i_{h\leftarrow c}\bigr|$ \\

    \bottomrule
  \end{tabular}
\end{table}

This leaves open the further normative question of what makes the underlying epistemic relations wrongful.
We now deploy the deficit schema in Eq.~\eqref{eq:deficit} to give a unified treatment of epistemic injustices in algorithmic settings. The next tables articulate this treatment at two complementary levels. Tables \ref{tab:ei_technical} and \ref{tab:ei_technical_forced} provide the quantitative operationalization of the schema, while Tables \ref{tab:ei_philosophical} and \ref{tab:ei_philosophical_forced} situate those formalizations in their corresponding epistemic mechanisms. Within the technical presentation, some injustices naturally appear as ideal--actual shortfalls, whereas others are more fundamentally relational, involving asymmetric flows or misallocations. Even in the latter cases, we recast the harm in deficit form in order to preserve a unified formal representation across contexts.

The next section maps these deficits to concrete algorithmic mechanisms across digitally-mediated domains.
\begin{table}
  \caption{\textbf{Technical representations of relational or flow-based epistemic injustices, recast in deficit form.} Arrow orientation indicates the direction of epistemic movement.}
  \label{tab:ei_technical_forced}
  \begin{tabular}{p{0.16\linewidth} p{0.31\linewidth} p{0.31\linewidth} p{0.12\linewidth}}
    \toprule
    Injustice & Ideal condition & Non-ideal condition & Deficit \\
    \midrule

    Epistemic exploitation \cite{berenstain2016epistemic} &
    Each agent contributes a fair share of epistemic labour $\ell_c$: $\ell_{a\leftarrow c} = \ell_c$. &
    Agent $a$ contributes
    $\ell_{a\leftarrow c} > \ell_c$ due to oppressive demand. &
    $\bigl|\ell_c - \ell_{a\leftarrow c}\bigr|$ \\

    Epistemic appropriation \cite{davis2018epistemic} &
    Agent $a$ receives the appropriate epistemic credit $\kappa_{a\to c}$: $\kappa_{a\to d} = \kappa_{a \to c}$. &
    Only $\kappa_{a\to d}$ is assigned to $a$, while remaining credit is diverted to dominant agents $d$:
    $\kappa_{a\to d} < \kappa_{a\to c}$. &
    $\bigl|\kappa_{a\to c} - \kappa_{a\to d}\bigr|$ \\

    Epistemic objectification
    \cite{haslanger2017objectivity,mcglynn2021epistemic} &
    Hearer $h$ fully recognizes speaker $s$'s epistemic agency $a_s$:
    $a_{h\leftarrow s} = a_s$. &
    Hearer $h$ treats $s$ as a source of information,
    recognizing less agency: $a_{h\leftarrow s} < a_s$. &
    $\bigl|a_s - a_{h\leftarrow s}\bigr|$ \\

    Hermeneutical marginalization \cite{fricker2007epistemic} &
    All agents have equal access to hermeneutical
    participation $p_c$: $p_{a\leftarrow c} = p_c$. &
    Agent $a$'s actual hermeneutical participation is restricted:
    $p_{a\leftarrow c} < p_c$. &
    $\bigl|p_c - p_{a\leftarrow c}\bigr|$ \\

    \bottomrule
  \end{tabular}
\end{table}
\begin{table}[]
  \caption{\textbf{Philosophical characterizations of epistemic injustices represented as direct ideal--actual shortfalls in Table \ref{tab:ei_technical}.}}
  \label{tab:ei_philosophical}
  \begin{tabular}{p{0.26\linewidth} p{0.68\linewidth}}
    \toprule
    Injustice & Mechanism \\
    \midrule

    Contributory injustice \cite{dotson2011tracking} &
    Marginalized knowers' interpretative contributions are dismissed/forced through dominant frameworks, so their situated knowledge fails to receive uptake. \\

    Formative epistemic injustice \cite{nikolaidis2021third} &
    Wrongs to a person's development as a knower, produced by social structures or institutions that limit the formation of epistemic capacities. \\

    Hermeneutical injustice \cite{fricker2007epistemic} &
    A person's sense-making capacity is undermined by identity-based prejudice. \\

    Prediscursive epistemic injury \cite{lobb2018prediscursive} &
    Structural conditions undermine a person's epistemic agency prior to participation in explicit testimonial exchanges (e.g., by undermining self-trust). \\

    Testimonial injustice
    \cite{fricker2007epistemic,quaresmini2025role} &
    A speaker’s credibility is downgraded by a hearer’s identity-based prejudice.\\

    Testimonial quieting \cite{dotson2011tracking} &
    The audience withholds uptake by failing to recognize the speaker as a knower. \\

    Testimonial smothering \cite{dotson2011tracking} &
    The speaker truncates or withholds testimony in anticipation of hostile uptake. \\

    Willful hermeneutical ignorance \cite{pohlhaus2017varieties} &
    Powerful agents choose not to know marginalized perspectives despite available evidence and interpretative tools. \\

    \bottomrule
  \end{tabular}
\end{table}

\begin{table}
  \caption{\textbf{Philosophical characterizations of relational or flow-based epistemic injustices recast in deficit form in Table \ref{tab:ei_technical_forced}.}}
  \label{tab:ei_philosophical_forced}
  \begin{tabular}{p{0.26\linewidth} p{0.68\linewidth}}
    \toprule
    Injustice & Mechanism \\
    \midrule

    Epistemic exploitation \cite{berenstain2016epistemic} &
    Marginalized knowers are systematically overburdened with epistemic
    labour (e.g., educating privileged agents about oppression), under
    conditions where this labour is uncompensated and often unsafe. \\

    Epistemic appropriation \cite{davis2018epistemic} &
    Dominant agents appropriate epistemic resources from marginalized communities, while credit and benefits bypass the original knowers. \\

    Epistemic objectification
    \cite{haslanger2017objectivity,mcglynn2021epistemic} &
    A knower is treated as a mere source or object of information rather
    than as an epistemic subject with their own perspective, reasons, and
    standing. \\

    Hermeneutical marginalization \cite{fricker2007epistemic} &
    Members of a social group are structurally excluded from participating
    in the collective hermeneutical practices that shape shared interpretative
    resources. \\

    \bottomrule
  \end{tabular}
\end{table}

\section{Algorithmic Instantiations of Epistemic Injustices}\label{sec:alg}
We map epistemic deficits to the socio-technical pipelines that can produce, 
stabilize or amplify them by tracing the lifecycle of algorithmic mediation. Epistemic deficits can be induced at multiple stages, that is, in representation and labeling (what categories exist, how content and users are described), ranking, recommendation and targeting (who is seen and by whom), moderation and enforcement (which speech is removed, down-ranked, or demonetized), scoring and reputation (who is treated as credible or risky by default). Across these stages, epistemic injustice can persist even when standard group-level fairness constraints are satisfied \cite{quaresmini2023qualification,quaresmini2025role,villa2025epistemic,VILLA2025151}, since they 
regulate parity in 
the quality of classification outputs. Here, 
the harms at issue concern system-mediated 
epistemic resources and 
participation.
These mechanisms manifest in domain-specific forms, including:
\begin{itemize}
    \item Content-ranking and search engines may systematically down-rank, mislabel, or filter marginalized users' interpretative contributions through dominant taxonomies \cite{noble2018algorithms,lau2022content,thach2024visible,lee2024people,lin2023trapped} (\emph{contributory injustice} \cite{pozzi2024machine});
    \item Adaptive learning platforms 
    can present systematically less challenging curricula or fewer exploratory resources 
    to students from marginalized groups \cite{o2017weapons,chakraborty2024ethical,boateng2025algorithmic,baker2022algorithmic,bird2025algorithms}, thereby limiting opportunities for epistemic development and constraining the formation of epistemic capacities over time \cite{raisa2024epistemic,babu2025epistemic,hummel2025ethical,tanchuk2025personalized} (\emph{formative epistemic injustice});
    \item Advertising and targeting systems may systematically exclude marginalized users from ads providing information needed to make sense of own experiences (e.g., campaigns about legal remedies, healthcare services, or specific forms of discrimination) \cite{baumann2024fairness,datta2018discrimination} (\emph{hermeneutical injustice} \cite{milano2025algorithmic});
    \item Pervasive scoring 
    systems can assign persistently low reliability to certain groups before any act of testimony (e.g., treating them as untrustworthy by default in fraud detection) \cite{eubanks2025automating,benjamin2019} (\emph{prediscursive epistemic injury});
    \item Social networks and decision-support systems incorporating user reports 
    may systematically down-rank 
    marginalized users in 
    recommendation, or decision pipelines \cite{quaresmini2025role,villa2025epistemic,williams2025social,casu2025demographic} (\emph{testimonial injustice});
    \item Platform moderation 
    systems may systematically fail to recognize some users as knowers, for example by disproportionately labelling their posts as low-quality or spam, thereby reducing the visibility and uptake of their contributions as epistemically relevant
    \cite{lee2024people,davidson2019racial,thach2024visible,habibi2024content,delmonaco2024you,tobi2024towards} (\emph{testimonial quieting} \cite{harb2025testimonial});
    \item Creators may self-censor to avoid demonetization or removal \cite{dergacheva2023we,are2022shadowban,zeng2022content,kingsley2022give,kumar2019algorithmic} (\emph{testimonial smothering} \cite{harb2025testimonial});
    \item Designers who refuse to extend datasets, despite evidence of systematic misrepresentation \cite{keyes2018misgendering,hamidi2018gender,barocas2016big,d2023data} can produce systems that embody a selective blindness to marginalized standpoints (\emph{willful hermeneutical ignorance}).
\end{itemize}

Despite this, mainstream algorithmic fairness rarely treats epistemic injustice as a target of measurement and evaluation, and related discussions often remain highly abstract \cite{edenberg2023epistemic,nihei2022epistemic,galasinski2023epistemic,hull2023dirty,hancox2021epistemic,rafanelli2022justice,symons2022epistemic,sardelli2022epistemic,kim2022democracy,medvecky2018fairness,jalali2020information}.
To investigate these phenomena, we therefore need quantitative tools that can capture and compare epistemic deficits.

\section{From Partial Constraints to a Global, Dynamic Evaluation}\label{sec:metrics}

Much of the algorithmic fairness literature treats fairness primarily as a question about the quality of classification, i.e., how a (typically supervised) classifier allocates benefits across individuals or groups 
\cite{elzayn2019fair,quaresmini2023qualification,kinchin2024voiceless,hosseini2025epistemic}. 
In these settings, fairness is operationalized through constraints on predictions, labels, or error rates
~\cite{pessach2022review,barocas2020fairness,mehrabi2021survey,verma2018fairness,narayanan2018translation}. More specifically, standard group fairness metrics are defined as relations among probabilities attached to model outputs. These metrics typically evaluate a fixed decision rule, that is, a static algorithm whose promotion rates are designed to satisfy a given parity condition across groups. For instance, equality of opportunity requires that the probability of receiving a positive decision 
be the same across groups 
among those individuals who qualify for the positive outcome.

Standard fairness evaluations of this kind are therefore snapshot-oriented. Indeed, they assess whether, at a given step, a fixed prediction rule applied to a fixed population satisfies the relevant constraint \cite{quaresmini2024data}. 
These metrics are highly useful for evaluating classifier performance and the fairness of decision rules \cite{lum2016statistical,quaresmini2024data,scantamburlo2025prediction}, typically through equality- or equity-based criteria. 
However, these approaches are not sufficient for our purposes. Indeed, rather than evaluating predictive performance or one-shot decision tasks, we want to assess the epistemic impact of ongoing platform-user interaction. This continuative assessment is particularly important as ranking, moderation, and feedback mechanisms dynamically shape who is heard and rendered intelligible over time, generating cumulative effects beyond isolated outputs as exemplified next. 

\indent\textit{Toy example: equality of opportunity in a platform feedback loop}.
Consider a professional social platform such as LinkedIn, where posts are ranked in users' feeds. Suppose the platform uses a predictive model to decide whether a post should receive high visibility. 
Let the promotion in the feed be the positive decision for posts qualified for high visibility. A standard fairness requirement 
is equality of opportunity 
across protected groups 
(e.g., men and women). Under this requirement, promotion rates should be equal among qualified posts. 
In a platform setting, however, promotion is rarely an isolated event. 
Visibility affects impressions and engagement, which in turn influence future rankings, creating a feedback loop.
Equality of opportunity can still be assessed at any given round, but this does not exhaust the evaluation. 
It is also important to understand how repeated ranking and engagement updates reshape the distribution of an epistemic resource.
In particular, here the epistemic resource is \emph{epistemic authority}, i.e., the standing agents acquire as credible and influential contributors through repeated exposure and uptake in the feed. The question, then, is not only whether promotion satisfies parity at one step, but also whether the feedback loop concentrates epistemic authority, produces lower-tail deprivation, or generates diverging epistemic standing within and across groups.

These
aspects are however not assessed by mainstream fairness metrics, that are mostly \emph{partial} constraints. Indeed, they test whether a relation holds across groups along a selected dimension, such as equal acceptance or false positive rates
\cite{pessach2022review,barocas2020fairness,mehrabi2021survey,verma2018fairness,narayanan2018translation} , but 
do not capture the full distribution across individuals. As a result, two systems may satisfy the same group-level parity constraint while differing in within-group dispersion, tail deprivation, or elite concentration. These aspects are particularly important to assess epistemic injustice, as marginalization operates beyond average group disadvantage,
but also through lower-tail deprivation, concentrated authority, and stratified access within and across groups. 
This motivates an explicit distributive perspective, connected to work on distributive justice 
\cite{hertweck2024s,kuppler2022fair,lundgard2020measuring,baumann2022distributive}. 
Building on \cite{fabris2022tackling}, predictive outcomes can often be read as allocative decisions, by treating positive predictions as a resource that the system \emph{assigns}. This highlights a broader point, namely, the content of a fairness analysis depends on how the relevant resource is specified in a given setting.
In social and information platforms, for example, the resource may be information itself, and fairness concerns how that epistemic good flows through a network \cite{quaresmini2023qualification}. 
Once predictions are treated as assignments, two evaluative questions arise. One concerns predictive quality, including error rates, calibration, and their parity across groups. The other concerns the distributive structure of the resulting allocations across individuals (e.g., whether they are top-heavy, stratified, or segregated). Standard predictive fairness metrics primarily track performance quantities. 
In our setting, evaluating epistemic injustice requires distribution-sensitive measures that can complement performance-oriented metrics and support longitudinal analysis.
Fairness indices are useful here because they provide explicitly \emph{global} summaries. They take as input the full distribution of a good or burden across agents and summarize its deviation from a benchmark, such as equality or a specified social weighting. This allows comparisons of distributional structure even when group-level constraints are satisfied. Because such indices are less common in ML-focused evaluations, their uptake as general-purpose fairness tools has remained limited. Existing connections include the Jain index in recommender systems \cite{rampisela2024evaluation}, ranking optimization with generalized Gini criteria \cite{do2022optimizing}, and unified perspectives on indices and fairness \cite{speicher2018unified}. However, these approaches have not been systematically integrated into the mainstream debate.

Introducing fairness indices into algorithmic fairness serves three purposes. First, it provides a common language for comparing distributional structure across systems, resources, and domains. Second, many indices are parameterized, which makes explicit how sensitive an evaluation is to the worst-off, the best off, or overall dispersion. Third, they yield scalar summaries that can be tracked over time and integrated into optimization, simulation, and ML pipelines.
To apply these indices in the epistemic setting, we must specify the epistemic quantity $\nu_i$ whose distribution is being evaluated, either an epistemic resource
directly 
or an output-induced measure of epistemic opportunity.
 We now make this explicit by characterizing epistemic goods
and their operational proxies in algorithmic systems.

\section{The Two Facets of Epistemic Auditing}\label{sec:indices}
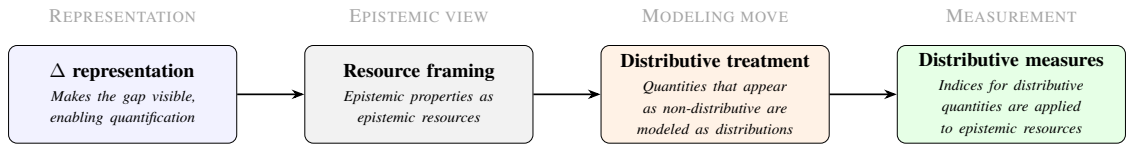
\begin{figure}[]
\centering
\resizebox{0.9\linewidth}{!}{%
\begin{tikzpicture}[
  node distance=0.9cm and 1.1cm,
  auto,
  base/.style={rectangle, draw, text width=3.5cm, align=center,
      rounded corners, minimum height=1.6cm, font=\small},
  nA/.style={base, fill=blue!5},
  nB/.style={base, fill=gray!10},
  nC/.style={base, fill=orange!10},
  nD/.style={base, fill=green!10},
  line/.style={draw, -{Stealth[scale=0.9]}, thick},
  note/.style={font=\scshape\footnotesize, color=gray!70}
]

\node[nA] (delta) {
  \textbf{$\Delta$ representation}\par\vspace{0.3ex}
  {\scriptsize\itshape Makes the gap visible, enabling quantification}
};

\node[nB, right=of delta] (resource) {
  \textbf{Resource framing}\par\vspace{0.3ex}
  {\scriptsize\itshape Epistemic properties as epistemic resources}
};

\node[nC, right=of resource] (dist) {
  \textbf{Distributive treatment}\par\vspace{0.3ex}
  {\scriptsize\itshape Quantities that appear as non-distributive are modeled as distributions}
};

\node[nD, right=of dist] (measures) {
  \textbf{Distributive measures}\par\vspace{0.3ex}
  {\scriptsize\itshape Indices for distributive quantities are applied to epistemic resources}
};

\node[above=0.25cm of delta, note] {Representation};
\node[above=0.25cm of resource, note] {Epistemic view};
\node[above=0.25cm of dist, note] {Modeling move};
\node[above=0.25cm of measures, note] {Measurement};

\draw[line] (delta) -- (resource);
\draw[line] (resource) -- (dist);
\draw[line] (dist) -- (measures);

\end{tikzpicture}%
}
\caption{\textbf{Conceptual pipeline underlying our framework.} The figure summarizes the diagnostic structure of the paper.
}
\label{fig:reasoning-pipeline}
\end{figure}

Starting our deficit-based formalization of epistemic injustice, we now proceed providing a set of indices allowing us to quantify the deficit distribution
across individuals and groups, toward completing the description of our diagnostic pipeline shown in Figure \ref{fig:reasoning-pipeline}.
Classical fairness and inequality indices were developed to summarize how unevenly a socially salient resource is distributed. In what follows, we introduce and repurpose them to measure the deficit terms associated with epistemic injustices. This yields a quantitative handle on epistemic harms that can persist even when conventional fairness constraints are satisfied.
Distributive analyses summarize how unevenly a good (or burden) is distributed by mapping a distribution to a scalar deviation from a normative benchmark. In quantitative work, this is often captured by \emph{fairness indices}, 
a family of measures with roots in welfare economics and broad uptake across policy and social science. In our framework, the same index 
supports two distinct forms of evaluation, depending on 
what quantity 
is being distributed. This distinction aligns with classic economic accounts on aversion to unequal outcomes and relative disadvantage \cite{fehr1999theory,bolton2006inequality,engelmann2004inequality}. On this view, unequal allocations matter not only in absolute terms, but also because agents may be sensitive to interpersonal asymmetries and relative standing within a distribution. In our framework, this broader insight motivates a distinction between inequalities in epistemic resources themselves and inequities in the epistemic opportunities mediated by system outputs.
In the \emph{resource inequality} case, the 
quantity of interest is a direct proxy for an epistemic resource (or burden) for each agent $i \in \mathcal{V}$:
\begin{equation}\label{eq:resource}
\nu_i \;=\; \rho_i \;\ge\; 0,
\end{equation}
where $\rho_i$ represents goods such as credibility standing, interpretative access, or epistemic authority. For example, on a professional platform such as LinkedIn, $\rho_i$ may represent the epistemic authority an agent has accumulated through visibility or uptake of their contributions. Here, the analysis asks how unevenly the system distributes or concentrates epistemic resources across individuals or groups. In the \emph{capability/rights inequity} case, by contrast, the quantity of interest is not the resource directly, but the epistemically relevant consequence induced by a system output. 
Let $y_i$ denote a system output for agent $i$ in context $\mathcal{C}$, and define:
\begin{equation} \label{eq:ouput_vi}
\nu_i \;=\; \phi(y_i, x_i, \mathcal{C}) \;\ge\; 0,
\end{equation}
where $\phi$ maps the output $y_i$ and agent features $x_i$ to an epistemically meaningful magnitude, such as the probability of being heard or expected epistemic opportunity. For example, if $y_i$ is the ranking position assigned to a post in a LinkedIn-style feed, $\phi(y_i, x_i, \mathcal{C})$ may represent the resulting probability that the post is seen, engaged with, and taken up as epistemically relevant. Here, the analysis asks how system decisions translate into unequal downstream opportunities, access conditions, or constraints.
Once the meaning of $\nu_i$ is fixed, the value of $\nu_i$ can vary over time based on platform-user interactions and, hence, any fairness index has to be assessed over time. In providing the definition of the fairness indices considered in this work, we nonetheless neglect this time dependence for notational simplicity. Moreover, note that all the
considered indices are agnostic to the nature of $\nu_i$, and hence can be applied both when $\nu_i$ is intended in the sense of Eq.~\eqref{eq:resource} or in that of Eq.~\eqref{eq:ouput_vi}.

What counts as an epistemic resource is context-dependent. In our framework, the term does not refer to a single homogeneous good, but to any epistemically salient quantity whose allocation, access, or uptake is mediated by the system. Depending on the setting, this may include interpretative resources, credibility standing, epistemic authority, uptake, or epistemic labour. Accordingly, $\nu_i$ should be understood as a measurable proxy for that quantity in context, rather than as a direct measure of some epistemic value. Operationally, specifying $\nu_i$ involves a modeling choice that should be performed cautiously. The relevant aspect of the epistemic good has to be identified for the application, and an observable proxy has to be selected to track it. In input-oriented analyses, $\nu_i$ may represent a baseline resource or burden already possessed by the agent. In output-oriented analyses, $\nu_i$ may instead be constructed from system outputs as an induced opportunity or entitlement. 
Note that $\nu_i$ does not capture the full epistemic wrong, but a distribution-sensitive aspect that can be compared across agents, groups and system configurations.

\indent\textit{Illustrative parametrizations}.
Let us first consider a case in which a platform, e.g., a recommender or advertising system, iteratively interact with users. Therefore, the platform serves a feed, the user interacts with it and, accordingly, their profile is updated, and future content adapts accordingly. In this case, in a resource-oriented analysis, $\nu_i$ may be proxied by the share of interpretative content shown to user $i$, or any binary indicator of whether user $i$ is exposed to resources relevant to understand a given condition. Meanwhile, in an output-oriented analysis, $\nu_i$ can be represented by the credibility standing of the user subsequent to the interaction with a platform (and eventually other users), namely the degree to which a user is treated as reliable or trustworthy by the platform after such an interaction. As another example, consider a platform that suggests a user a set of possible rides, ranked by distance, location, traffic, or expected tip (e.g., Uber, Lyft, or autonomous taxi services). The user picks a ride and rates it, quantitatively assess the quality of the ride (e.g., in terms of comfort and punctuality). The driver is then given
a score by the platform, that shapes how future suggestions and hence ride assignment will be made. In this case, in resource-oriented analysis, $\nu_i$ might be proxied by the rider score itself, here interpreted as a measure of epistemic standing. Meanwhile, in an output-oriented analysis, $\nu_i$ might be a consultative counter of high-value rides received by driver $i$.


\section{Classical Indices and Their Translation in an
Epistemic Framework}\label{sec:epistemic}
We now introduce the indices proposed in this work to assess epistemic injustice and their translation into the epistemic framework. Note that they provide global summaries of the distribution of $\nu_i$, but they can be applied at different scales, such as over individuals to measure dispersion (e.g., concentration of authority), or on group-aggregated quantities to measure disparity (e.g., identity-based marginalization).
Let $\{\nu_i\}_{i=1}^N$ be nonnegative utilities with mean $\bar{\nu}=\tfrac{1}{N}\sum_i \nu_i$.

\indent\textbf{Dispersion- and concentration-based indices.} Dispersion indices summarize the uneven distribution of
a nonnegative quantity 
across a population. The \emph{Jain index} ($J$) \cite{jain1984quantitative} measures the \emph{evenness} 
of allocations $\nu_i$ 
across individuals, ranging from $1/N$ (maximal inequality) to $1$ (perfect evenness):
\begin{equation} \label{eq:jain}
J \;=\; \frac{\left( \sum_{i=1}^{N} \nu_i \right)^2}{N \cdot \sum_{i=1}^{N} \nu_i^2}.
\end{equation}
Because $J$ depends on the ratio between the first and second moments of $\{\nu_i\}$, it is 
sensitive to extreme values. A 
few very large $\nu_i$ can substantially lower $J$, even 
when most allocations are similar.
The \emph{Epistemic Jain index} (evenness of epistemic access and standing) captures whether epistemic standing or access is broadly shared across individuals, rather than concentrated among a few. 
In an input-oriented evaluation, low $J^{\mathrm{epi}}$ indicates an epistemic environment in which a substantial fraction of agents remain effectively excluded, even when standard group-level fairness is satisfied.
In an output-oriented evaluation, $J^{\mathrm{epi}}$ measures evenness in effective epistemic opportunity induced by the system and flags uneven participation conditions.
The \emph{Gini index} ($G$) \cite{gini1912variabilita} measures \emph{concentration} via average pairwise disparity:
\begin{equation}\label{eq:gini}
G \;=\; \frac{1}{2N^{2}\bar{\nu}}\sum_{i=1}^{N}\sum_{j=1}^{N}\bigl|\nu_i-\nu_j\bigr|.
\end{equation}
Equivalently, $G$ is the mean absolute difference between two randomly chosen individuals' allocations, normalized by $2\bar{\nu}$. Thus $G=0$ under perfect equality and increases as the quantity becomes more concentrated among fewer individuals. 
The \emph{Epistemic Gini index} (epistemic hierarchy and concentration) captures 
how strongly an epistemic quantity is stratified across individuals. This matters for hierarchy-oriented harms. A small subset of agents may accumulate epistemic capital, hermeneutical access, or developmental opportunity, leaving others persistently disadvantaged.
In an input-oriented evaluation, high $G^{\mathrm{epi}}$ indicates a stratified epistemic economy in which baseline standing or access is highly unequal. In an output-oriented view, high values indicate that the system 
concentrates epistemic opportunity.
The \emph{Hoover} or \textit{Robin Hood index} ($H$) \cite{hoover1936measurement} measures the share of the total amount that must be redistributed from individuals above the mean to those below it, to achieve equality:
\begin{equation} \label{eq:hoover}
H \;=\; \frac{1}{2N\bar{\nu}}\sum_{i=1}^{N}\bigl|\nu_i-\bar{\nu}\bigr|.
\end{equation}
A value of $H=0$ indicates perfect equality. The index increases with concentration and reaches its maximum value $H=1-\tfrac{1}{N}$ when a single individual holds all of the resources.
The \emph{Epistemic Hoover index} (distance to equal epistemic standing) 
measures how much epistemic standing, access, or opportunity has to be redistributed to 
remove inequality in epistemic participation.
For inputs, high $H^{\mathrm{epi}}$ means that equal standing or access would require redistributing a substantial share of epistemic resources. For outputs, it indicates a large redistribution to equalize 
epistemic opportunities, even when mean opportunity is high.
The \emph{Generalized entropy} ($GE$) family \cite{cowell2000measurement, theil1967economics}  quantifies dispersion in nonnegative distributions. $GE$ measures compare each allocation $\nu_i$ to the population mean $\bar{\nu}$, yielding $GE(\alpha)=0$ under perfect equality and larger values as the distribution becomes more uneven.
It is defined as:
\begin{equation}\label{eq:generalized_entropy_app}
GE(\alpha) =
\begin{cases}
\displaystyle 
\frac{1}{\alpha(\alpha - 1)}
\left( \frac{1}{N} \sum_{i=1}^{N}
\left[ \left( \frac{\nu_i}{\bar{\nu}} \right)^{\alpha} - 1 \right] \right),
& \alpha \ne 0,1, \\[0.7ex]
\displaystyle 
-\frac{1}{N} \sum_{i=1}^{N}
\ln\!\left( \frac{\nu_i}{\bar{\nu}} \right),
& \alpha = 0 \text{ (Theil $L$)}, \\[0.7ex]
\displaystyle 
\frac{1}{N} \sum_{i=1}^{N}
\left( \frac{\nu_i}{\bar{\nu}} \,
\ln\!\left( \frac{\nu_i}{\bar{\nu}} \right) \right),
& \alpha = 1  \text{ (Theil $T$)}.
\end{cases}
\end{equation}
The parameter $\alpha \in \mathbb{R}$ tunes tail sensitivity. Smaller $\alpha$ places more weight on disparities among lower allocations (with $\alpha<0$ being especially sensitive to the bottom tail), while larger $\alpha$ emphasizes disparities among higher allocations (notably for $\alpha>1$).
\emph{Generalized epistemic entropy family} (tail-sensitive epistemic inequality) 
distinguishes whether epistemic unfairness is driven by exclusionary lower tails (many agents with near-zero access or uptake) or by extreme top concentration (a small subset captures disproportionate advantage). 
For inputs, it summarizes tail-sensitive inequality in baseline epistemic resources. For outputs, it measures tail-sensitive inequality in epistemic opportunity.
The \textit{Atkinson index} ($A$) \cite{atkinson1970measurement} measures inequality in the distribution of a nonnegative resource across a population. It can be read as the share of mean allocation $\bar{\nu}$ society would forgo in order to achieve equality, given an inequality aversion parameter $\varepsilon \ge 0$.
Larger values of $\varepsilon$ place greater weight on the well-being of individuals with lower allocations. $A(\varepsilon) = 0$ indicates perfect equality, and larger values greater inequality.
\begin{equation}\label{eq:atkinson_app}
A(\varepsilon) =
\begin{cases}
1 - \dfrac{1}{\bar{\nu}}
\left( \dfrac{1}{N} \sum_{i=1}^{N}
\nu_i^{1 - \varepsilon} \right)^{\frac{1}{1 - \varepsilon}},
 & \varepsilon \neq 1, \\[0.7ex]
1 - \dfrac{1}{\bar{\nu}}\left( \dfrac{1}{N} \prod_{i=1}^{N} \nu_i\right)^{\frac{1}{N}}, & \varepsilon = 1.
\end{cases}
\end{equation}

The \emph{Epistemic Atkinson index} (prioritarian sensitivity to low standing) 
implements a prioritarian view. Inequality counts more when it disadvantages agents with low epistemic standing or access. 
For inputs, high $A^{\mathrm{epi}}(\varepsilon)$ indicates a baseline distribution that is costly under priority to low-standing agents. For outputs, high $A^{\mathrm{epi}}$ indicates 
lower-tail disadvantages, even when total opportunity is high.

\indent\textbf{Separation-based indices.} Separation indices capture how strongly two groups are unevenly distributed across categories or positions, that is, whether members of different groups tend to concentrate in distinct segments of a distribution rather than being proportionally represented within each unit. Let units be indexed by $k\in{1,\dots,n}$. Denote by $a_k$ and $b_k$ the counts of group $A$ and group $B$ in unit $k$, and by $A=\sum_k a_k$ and $B=\sum_k b_k$ their total population.
The \emph{Dissimilarity index} ($D$) \cite{WHITE2005403} measures the degree of separation between the two groups as
\begin{equation} \label{eq:dissimilarity}
D = \frac{1}{2}\sum_{k=1}^{n}
\left|
\frac{a_k}{A}-\frac{b_k}{B}
\right|.
\end{equation}
$D$ ranges from $0$ (perfect mixing, where each unit mirrors the overall group proportions) to $1$ (complete separation, where units are occupied exclusively by one group). It can be interpreted as the fraction of either group that would need to be reassigned across units in order to achieve an identical distribution.
The \emph{Epistemic Dissimilarity index} (separation into epistemic environments) targets structural harms in the allocation of epistemic resources and opportunities, since separation across environments can sustain unequal epistemic opportunity despite similar within-unit treatment. 
For inputs, $D^{\mathrm{epi}}$ can be computed from participation patterns across environments (e.g., who appears in feeds). For outputs, from induced access or opportunity (e.g., who is sent to environments with greater epistemic opportunity). It thus captures marginalization through structured separation, not only individual-level disparity.


\indent\textbf{Percentile ratios and tail measures.} The \textit{Palma ratio} ($P$) \cite{palma2011homogeneous} is defined as the ratio of the income share held by the top $10\%$ (the richest) to that held and bottom $40\%$ (the poorest) of the population.
\begin{equation}
P \;=\;
\frac{\sum_{i \in \text{top }10\%} \nu_i}{\sum_{i \in \text{bottom }40\%} \nu_i}.
\end{equation}
Higher values indicate highest concentration at the top.
When $P<1$ ($P>1$), the bottom $40\%$ holds a larger (smaller) share of total resources than the top $10\%$.
The \emph{Epistemic Palma ratio} (elite capture versus broad-base access) 
detects elite capture of epistemic resources. Indeed, a small epistemic elite may accrue visibility, uptake, or credibility. By contrast, the broad base collectively receives little.
From an input-oriented view, $P^{\mathrm{epi}}$ compares baseline epistemic capital in the top $10\%$ to that in the bottom $40\%$. From an output-oriented perspective, it compares realized epistemic opportunity 
accrued by the top $10\%$ to the one by the bottom $40\%$. It can reveal top accumulation even when group-level parity constraints are satisfied.
The \emph{Quintile share ratio} ($S80/S20$) \cite{drezner2014quintile}, compares the total resources held by the top income quintile (the richest $20\%$) to those held by the bottom income quintile (the poorest $20\%$):
\begin{equation}
S80/S20= \frac{\sum_{i \in \text{top }20\%} \nu_i}{\sum_{i \in \text{bottom }20\%} \nu_i}.
\end{equation}
Higher values indicate stronger concentration at the top. 
The \emph{Epistemic Quintile share ratio} (top--bottom epistemic opportunity) compares the total held by the top $20\%$ to that held by the bottom one. It 
captures the top-bottom gap in epistemic standing or access. 
It compares top and bottom quintiles in epistemic resources (inputs) or opportunity (outputs), highlighting top-heavy participation and uptake even when within-group performance is similar.

\indent\textit{Example (epistemic capital).}
Let $v_i\ge 0$ denote agent $i$'s epistemic capital (e.g., a proxy for credibility standing or authority).
For an input-oriented evaluation, 
the epistemic Jain (Eq.\eqref{eq:jain}) and Gini  (Eq.\eqref{eq:gini}) indices summarize inclusion and hierarchy in baseline epistemic capital. In the ride-assignment example introduced above, this corresponds to evaluating how evenly drivers' platform scores (used as a proxy for epistemic standing within the system) are distributed, and whether such standing becomes concentrated among a limited subset of drivers.
Low $J^{\mathrm{epi}}$ indicates that many agents have near-zero epistemic capital (exclusion, e.g., many drivers have very low platform scores and are therefore weakly recognized as reliable by the system), while high $G^{\mathrm{epi}}$ indicates stratification in standing (hierarchy, e.g., credibility is strongly concentrated among a smaller set of highly scored drivers), even if only a minority is fully excluded.
For an output-oriented analysis,
let $\nu_i$ denote output-induced epistemic opportunity (e.g., expected uptake or exposure that contributes to realized epistemic capital over time). The same indices summarize how evenly the system distributes epistemic opportunity ($J$) and how strongly that opportunity is concentrated ($G$). In the ride-assignment setting, $\nu_i$ may instead capture downstream opportunities such as the number of high-value rides, premium assignments, or repeated access to profitable requests allocated to driver $i$.
Similarly, when considering the epistemic translation of indices in Eqs.~\eqref{eq:hoover}-\eqref{eq:generalized_entropy_app} and \eqref{eq:atkinson_app}, high $H^{\mathrm{epi}}$ indicates a large distance to equalize epistemic capital (repair cost, e.g., the extent of redistribution needed to equalize either driver scores or access to valuable rides). $GE^{\mathrm{epi}}(\alpha)$ makes inequality diagnosis tail-sensitive, with lower $\alpha$ emphasizing exclusion among low-capital agents (e.g., low-scored or rarely assigned drivers) and higher $\alpha$ emphazising concentration at the upper-tail, namely among drivers who repeatedly receive the most advantageous assignments. $A^{\mathrm{epi}}(\varepsilon)$ instead encodes a prioritarian sensitivity to low-capital agents that grows with $\varepsilon$, thus placing increasing normative weight on drivers who remain persistently at the bottom of the score or opportunity distribution.
If the platform partitions interactions into units or environments $k\in\{1,\dots,n\}$ (e.g., feeds, communities, categories), we can compute an epistemic dissimilarity score $D^{\mathrm{epi}}$ to assess whether protected groups are systematically separated into clusters with different epistemic affordances (e.g., routed away from high-uptake spaces), even when within-environment treatment does not differ by group. This could capture whether some groups of drivers are systematically routed toward less profitable zones, time slots, or ride categories, even if assignments are locally comparable within each environment. 
Finally $P^{\mathrm{epi}}$ (Palma) and $S80/S20^{\mathrm{epi}}$ ratios provide coarse summaries of elite capture and top-bottom gaps in epistemic capital, indicating whether high platform scores or valuable ride opportunities are disproportionately concentrated among a small subset of drivers relative to the least advantaged ones.

We remark that the same index can be used in both cases, but its normative interpretation differs. Resource-oriented analyses evaluate how epistemic goods or burdens are distributed, whereas output-oriented analyses diagnose how system decisions mediate access to epistemically relevant opportunities. 
The proposed indices offer a diagnostic account of how unfairness arises and evolves across resource and outcome dimensions. Beyond diagnosis, they can also support fairness-oriented intervention, by linking observed values to corrective actions and by enabling a dynamic assessment of whether those actions reduce unfairness over time.

\section{Simulation Example} \label{sec:sim_example}
We now illustrate how the proposed fairness indices can be used to evaluate the evolution of epistemic quantities on a digital platform. The platform is modeled as a recommender system that reshapes interactions among agents by modifying users' influence on each other. 
\begin{figure}[]
    \centering
    \begin{subfigure}{0.4\linewidth}
        \centering
        \includegraphics[width=0.6\linewidth, angle=180]{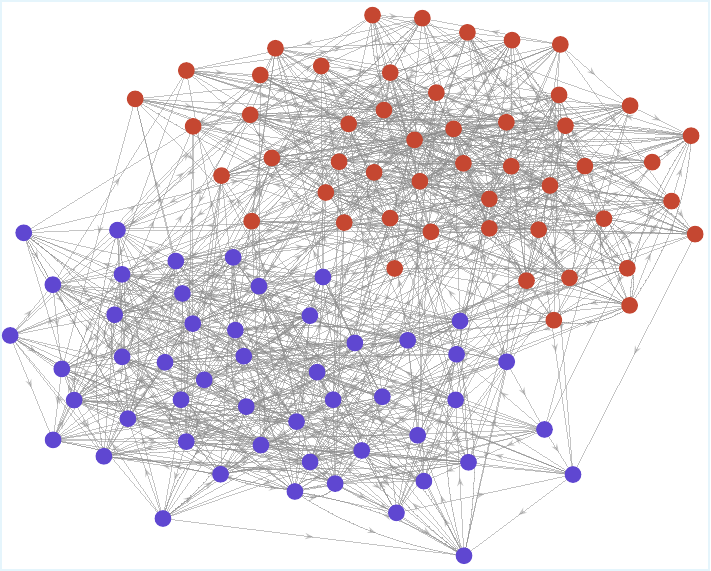}
        \caption{Network directed graph.}
        \label{fig:network}
    \end{subfigure}
    \hspace{0.03\linewidth}
    \begin{subfigure}{0.4\linewidth}
        \centering
        \resizebox{0.4\linewidth}{!}{%
            \input{beta_distr_groupAB}
        }
        \caption{Probability Density Function of $x_0$ among groups.}
        \label{fig:pdf}
    \end{subfigure}
    \caption{\textbf{Simulation framework.} Network structure and distribution of initial opinions for the two groups. Group A information is depicted in red, while Group B is indicated in blue.}
    \label{fig:netw_pdf}
\end{figure}
Operationally, this captures mechanisms such as visibility boosting or shadowbanning, through which the platform amplifies or reduces the exposure of specific users, therefore redistributing \emph{epistemic resources}.
To analyze how these interventions propagate over time, we embed the recommendation mechanism within an opinion dynamics framework, following \cite{kuhne2025optimizing}. This allows us to evaluate fairness from two complementary perspectives, namely, resource inequality (Eq. \eqref{eq:resource}) 
through the distribution of incoming attention, and outcome inequity (Eq. \eqref{eq:ouput_vi}) 
through the distribution of realized opinions.
The fairness indices introduced in Section \ref{sec:epistemic} are then applied to quantify disparities in both dimensions over the considered time horizon of $50$ instants.
We model the social network as a directed, weighted graph in which nodes represent agents and edges capture influence relations. The weight associated with a directed edge measures the strength of influence and, as in~\cite{kuhne2025optimizing}, the recommender system acts by modifying these weights over time.
The 
network 
reported in Fig.~\ref{fig:network} 
is composed of $N=100$ agents divided into two same-size groups, generated using a stochastic block model with higher intra-group than inter-group connectivity. This 
allows to capture homophily, as initial opinions are polarized across groups following a skewed Beta distribution (Fig.~\ref{fig:pdf}).
We compare three kinds of interventions: (i) \textit{baseline} (the weight matrix remains fixed), (ii) \textit{targeted boost} (the platform periodically amplifies the outgoing weights of one group), (iii) \textit{random boost} (amplification is applied to a randomly selected subset of $N/2$ agents).
At intervention times, the platform amplifies selected influence links and then renormalizes the network to preserve its structural properties (see Appendix~\ref{sec:sim_details} for simulation details). 

\begin{figure}[]
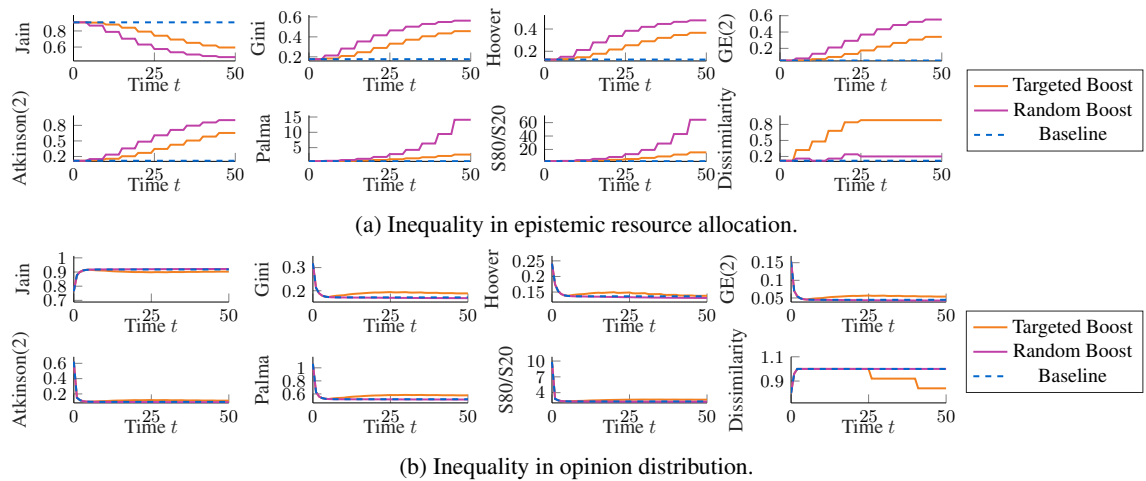

    \centering    
    \begin{subfigure}{\linewidth}
        \centering
        \resizebox{0.92\linewidth}{!}{%
            \input{input_indici_3}
        }
        \caption{Inequality in epistemic resource allocation.}
        \label{fig:input}
    \end{subfigure}   
    \begin{subfigure}{\linewidth}
        \centering
        \resizebox{0.92\linewidth}{!}{%
            \input{output_indici_3}
        }
        \caption{Inequality in opinion distribution.}
        \label{fig:output}
    \end{subfigure}    
    \caption{\textbf{Intervention effects on fairness indices}. Analysis on input and output quantities comparing the three presented scenarios.}
    \label{fig:indici}
\end{figure}

\noindent 
Fig.~\ref{fig:input} shows that both interventions increase inequality in epistemic resources distribution relative to the baseline, for all indices but the Dissimilarity. 
In the targeted boost case, the increase in overall inequality is more moderate,
 largely because amplification interacts with subsequent row-normalization. Since agents are more densely connected within their own group, 
scaling one group's columns mainly reallocates weight among already strongly connected nodes. After renormalization, attention redistribution remains relatively contained within the group's internal structure, so global dispersion of epistemic resources increases only comparatively modestly.
By contrast, random amplification spreads attention more broadly, producing a stronger increase in global inequality. However, group-based segregation follows a different pattern. The Dissimilarity index rises substantially under the targeted intervention, showing that epistemic advantage becomes aligned with group identity. Random amplification, while increasing overall dispersion, does not systematically reinforce group-level separation. This contrast between dispersion- and separation-based measures highlights the value of combining fairness metrics, since a policy may generate moderate global inequality while simultaneously intensifying identity-based epistemic segregation.

Concerning the output (Fig.~\ref{fig:output}),
the targeted boost increases dispersion in opinions, indicating 
that repeated within-group amplification translates into greater polarization of beliefs, while the random boost generates 
dynamics that remain close to the baseline for all indices. For the Atkinson and $S80/S20$ indices, the trajectories appear visually close near the end of the horizon, but their values confirm the same underlying pattern.
In particular, Atkinson reaches $0.112$ under the targeted boost, compared with $0.092$ under random amplification and $0.096$ in the baseline. Similarly, $S80/S20$ reaches $2.668$ under the targeted intervention, versus $2.287$ under random amplification and $2.330$ in the baseline. Although random intervention alters structural resource allocation, its effects on opinions are attenuated during the updating process.
However, the Dissimilarity index decreases under the targeted boost. Thus, although opinions become more dispersed overall, amplifying one group's outgoing influence shifts both groups toward similar opinions, reducing inter-group separation.

\section{Conclusions}\label{sec:conc}

Algorithmic systems increasingly mediate who is heard, believed, and understood. 
We argued that 
mainstream debates on algorithmic fairness largely treat fairness as 
predictive performance or outcome parity, and therefore lack an explicit epistemic account of what is at stake when systems govern credibility, uptake, and interpretative access. Yet many contemporary algorithmic infrastructures are intrinsically epistemic. Indeed, ranking, moderation, labeling and reputation systems 
not 
only allocate material benefits, but 
also shape the conditions under which agents 
participate as knowers. 
Epistemic harms, then, are not reducible to prediction errors or group-level parity constraints, but 
can emerge from 
system behavior at scale through the dynamics of algorithmic mediation. To make such harms measurable, we introduced a deficit-based formal template representing 
epistemic injustices as gaps between ideal and effective epistemic conditions, and mapped them to concrete stages of algorithmic mediation.
Building on this, we proposed fairness and inequality indices as global, distribution-sensitive tools for tracking epistemic harms over time. A central contribution is the distinction between two evaluation stances we draw in Section \ref{sec:indices}, namely, resource inequality, where indices are applied directly to epistemic goods and burdens, and capability/rights inequity, where indices are applied to output-induced epistemic opportunities. This separation clarifies how the same family of indices can support evaluation both before and after algorithmic mediation, and why purely predictive notions of fairness can be satisfied while epistemic participation remains structurally distorted. We then provided an epistemic translation of canonical indices, showing how they capture complementary signatures of epistemic unfairness.
The simulation study shows the practical value of the framework in a dynamic platform setting. Repeated interventions affect epistemic resource allocation and outcomes differently, supporting our broader claim that epistemic unfairness is a dynamic property of algorithmic mediation rather than a static parity condition.
Applying these measures requires operational proxies for epistemic resources (e.g., credibility standing, uptake, interpretative access, epistemic labour) and principled mappings from system outputs to epistemically meaningful quantities. While these choices are necessarily domain-specific, making them explicit provides a disciplined way to connect social-epistemological theories to measurable system properties. More importantly, it enables a form of evaluation that is crucial for systems whose primary effects are epistemic. Without tracking how systems distribute epistemic standing and access, fairness assessments risk optimizing for parity in outputs while leaving intact, or even amplifying, epistemic marginalization.
Several limitations and extensions are immediate. First, our deficit formalization is a measurement stance and does not fully capture the normative structure of epistemic wrongs. Future work should further articulate how deficit measures relate to the underlying wrongful epistemic relations, including deficit-excess dynamics and interactions across multiple epistemic goods. econd, the framework depends on context-sensitive proxies for epistemic goods, burdens, and opportunities. Proxy validity is therefore a central challenge. Indeed, if the chosen measure tracks the relevant epistemic quantity poorly, the resulting analysis may distort or obscure the very harm it is meant to diagnose.
Third, the indices 
are global summaries. Integrating them with intervention design and mechanism-level diagnostics remains an important direction. Finally, extending the framework to multigroup and intersectional settings, and to explicitly dynamic models of feedback and learning, would strengthen its applicability to real-world platforms.
Despite these limitations, our framework provides a formal bridge to include 
forms of epistemic injustice in algorithmic fairness evaluations. By treating epistemic harms as distributional distortions in epistemic goods and in 
epistemic opportunity, the framework complements partial fairness constraints and supports longitudinal auditing of epistemic participation, credibility, and interpretative access in contemporary algorithmic environments.

\bibliography{sample-base}
\bibliographystyle{ACM-Reference-Format}


\appendix
\section{Simulation Details} \label{sec:sim_details}

\paragraph{Network structure}
The network structure is generated using a stochastic block model where agents are more likely to connect within their own group than across groups. Formally, we set that, for each ordered pair of agents $(i,j)$, $i \ne j$, a directed edge from $i$ to $j$ is formed with probability
$$
\mathbb{P}(A_{ij}=1)=
\begin{cases}
p_{\text{intra}} = 0.18 & \text{if } i \text{ and } j \text{ belong to the same group}, \\
p_{\text{inter}} = 0.04 & \text{otherwise},
\end{cases}
$$
where $A_{ij}$ denotes the $(i,j)$-th entry of the adjacency matrix $A \in \{0,1\}^{N \times N}$. In particular, $A_{ij}=1$ if there exists a directed edge from agent, $0$ otherwise. 
Let $W(t) \in \mathbb{R}^{N \times N}$ denote the row-stochastic influence matrix at time $t$, where $w_{ij}(t)$ represents the weight (attention) that agent $i$ assigns to agent $j$. 
The elements of the initial influence matrix $W_0$ are drawn independently from a uniform distribution on $(0.5,1.5)$. The resulting matrix is row-normalized to obtain a row-stochastic influence matrix $W_0$. If an agent has no outgoing links, its row is replaced by a uniform distribution before normalization.
\paragraph{Initial opinions}
Each agent $i$ holds an initial opinion $x_{0,i} \in [0,1]$.  
Opinions are group-dependent and drawn from skewed Beta distributions to represent homophily (Fig.~\ref{fig:pdf}): 
$x_{0,i} \sim \mathrm{Beta}(1.4,5.0)$ if $i \in $ group A, and
$x_{0,i} = 1 - Z_i,  Z_i \sim \mathrm{Beta}(1.4,5.0)$ if $i \in$ group B.
\paragraph{Opinion dynamics model}
To model how opinions evolve over time, we employ the opinion dynamics theory. The considered opinion dynamics model captures the update of agents' opinions as the average of neighbors' opinions weighted by the influence matrix while retaining a fixed attachment to their initial belief.
As in~\cite{kuhne2025optimizing}, we employ the Friedkin--Johnsen (FJ) model~\cite{friedkin1990social}, according to which opinions (stacked in the vector state $x(t)$) evolve as
$$
x(t+1)=\Lambda x_0 + (I-\Lambda)W(t)x(t).
$$
Here, $I$ is the identity matrix and $\Lambda = \mathrm{diag}(\lambda_1,\dots,\lambda_N)$ represent the stubbornness matrix, where each agent $i$ stubbornness $\lambda_i$ is drawn independently from $\mathcal{U}(0.2,0.5)$.
\paragraph{Intervention mechanisms and measured quantities}
We consider three scenarios:
\begin{itemize}
    \item {baseline}: no intervention; $W(t)=W_0$ for all $t$.
    \item {targeted boost}: amplification of outgoing influence from all agents in Group A.
    \item {random boost}: amplification of outgoing influence from a fixed random subset of $N/2$ agents.
\end{itemize}

Interventions occur periodically over a horizon of $T=50$ steps, every $\lfloor T/10 \rfloor$ time instants.
At intervention times, the columns corresponding to the targeted agents are scaled by a factor $(1+\gamma)$ with $\gamma=0.5$.
The matrix is then renormalized to preserve row-stochasticity.  
At each time step, we compute the indices on:
\begin{itemize}
    \item Incoming attention: $\nu^{in}_i(t)=\sum_{j=1}^N w_{ji}(t)$, $\forall i$, which measures epistemic resources allocated to agent $i$;
    \item Realized opinions $\nu^{out}_i(t)=x_i(t)$, $\forall i$, capturing the epistemic consequences of influence dynamics.
\end{itemize}
The indices introduced in Section \ref{sec:epistemic} are then used to quantify these evolving inequalities. The Generalized Entropy is computed for $\alpha=2$, the Atkinson index for $\varepsilon=2$, and for the Dissimilarity index, the population is partitioned into $n_{\text{bins}}=10$ quantile bins, so that $a_k$ and $b_k$ denote the number of Group A and Group B agents in bin $k$.

\end{document}